\documentclass[epj]{svjour}

\usepackage{amssymb}
\usepackage{graphicx}
\usepackage{dcolumn}
\usepackage{amsmath}

\begin{document}

\title{Magnetic ordering and non-Fermi-liquid behavior in the multichannel
Kondo-lattice model}

\author{V. Yu. Irkhin%
\thanks{Valentin.Irkhin@imp.uran.ru}%
}
\institute{Institute of Metal Physics, 620990 Ekaterinburg, Russia}


\abstract{
Scaling equations for the Kondo lattice in the paramagnetic and magnetically
ordered phases are derived to next-leading order with account of spin dynamics.
The results are applied to describe various mechanisms of the
non-Fermi-liquid (NFL) behavior in the multichannel Kondo-lattice model 
where a fixed point occurs in the weak-coupling region.
The corresponding temperature dependences of
electronic and magnetic properties are discussed. The model describes
naturally formation of a magnetic state with soft boson mode and small moment value. 
An important role of Van Hove singularities in the magnon spectral function is
demonstrated. The results are rather sensitive to the type of magnetic
ordering and space dimensionality, the conditions for NFL behavior being
more favorable in the antiferromagnetic and 2D cases.}


\maketitle

\section{Introduction}

A number of 4$f$- and 5$f$-compounds, including so-called Kondo lattices and
heavy-fermion systems, possess anomalois electronic properties, e.g., giant
value of $T$-linear electronic specific heat $C(T)$ and magnetic
susceptibility $\chi _{m}(T)$ \cite{Stewart}. Magnetism of such systems
demonstrates also unusual features, including formation of an antiferro- or
ferromagnetic  state with small ordered moment value.

A common explanation of the heavy-fermion behavior is based on the Komdo
effect.  Unlike the one-impurity situation, the competition between the
Kondo screening of regularly arranged magnetic moments and intersite
magnetic interactions has a great importance in the lattice case. As a
result, smearing of Kondo singularities occurs on the scale of the
characteristic spin-dynamics frequency $\overline{\omega }.$ At the same
time, $\overline{\omega }$ itself acquires renormalizations due to the Kondo
screening. A scaling consideration of this renormalization process in the $%
s-f$ exchange model \cite{IK} yields, depending on the values of bare
parameters, both the ``usual'' states (a
non-magnetic Kondo lattice or a magnet with weak Kondo contributions) and
the peculiar magnetic Kondo-lattice state.

A great experimental material has been obtained for $f$-systems
demonstrating so-called non-Fermi-liquid (NFL) behavior \cite{Stewart1l},
which have unusual logarithmic or power-law temperature dependences of
electron and magnetic properties,.e.g.., $\chi _{m}(T)\sim T^{-\zeta }$ ($%
\zeta <1$), $C(T)/T$ is proportional to $T^{-\zeta }$ or $-\ln T$, for the
resistivity $R(T)\sim T^{\mu }$ ($\mu <2$), etc. The NFL behavior is
observed in a number of rare-earth and actinide systems: not only in alloys
like U$_{x}$Y$_{1-x}$Pd$_{3}$, UPt$_{3-x}$Pd$_{x}$, UCu$_{5-x}$Pd$_{x}$, CeCu%
$_{6-x}$Au$_{x}$, U$_{x}$Th$_{1-x}$Be$_{13}$, but also in some
stoichiometric compounds, e.g., Ce$_{7}$Ni$_{3}$, CeCu$_{2}$Si$_{2},$CeNi$%
_{2}$Ge$_{2}$ \cite{Proc}. It can coexist with magnetic ordering and occur
even in ferromagnets \cite{Nikiforov}.

There are a number of theoretical mechanisms proposed to describe the NFL
state, both single-site and intersite effects being discussed. In
particular, proximity to magnetic quantum phase transitions \cite{Vojta1}
should be mentioned.

The NFL behavior in the $M$-channel Kondo model (especially in the large-$M$
limit) was extensively investigated in the one-impurity case \cite%
{Tsv,Col,Gan,Cox,Parcollet}. Physically, this behavior is connected with
overscreening of impurity spin by conduction electrons. The model permits a
consistent scaling investigation since the fixed point is within the
weak-coupling region (however, the marginal case $M=2$ has some
peculiarities). On the other hand, the lattice case is more difficult and
only special approaches, in particular for one-dimensional models \cite%
{Col,Ventura} and infinite space dimension \cite{Jarrell} were used. In the
present paper we start from the standard microscopic model of a periodical
Kondo lattice and treat the interplay of the on-site Kondo screening and
intersite exchange interactions within a scaling approach. We will
demonstrate that, besides the standard one-impurity NFL mechanism,``soft'' 
boson branches can be formed  during the renormalization process,  
the role of singularities in spin spectral function being important for  the NFL behavior. 

Earlier a similar consideration was performed in Refs.\cite{IK,731} where
the NFL behavior in $M=1$ and large-$M$ Kondo lattices was treated within a
simple approximation corresponding to one-loop scaling (in the pseudofermion
representation). This approach yields NFL behavior in the formal limit $M \rightarrow \infty$ (where the coupling constant is unrenormalized, which is similar to occurrence of a fixed point), but for any realistic $M$  the NFL regime is achieved only in a very narrow interval of bare coupling constant (near the critical value for magnetic quantum phase transition).
Thus this approximation is insufficient to describe consistently the NFL state. In
the present work we perform the next-leading scaling analysis which changes radically the situation.

In Sect.2 we write down the scaling equations in the one-impurity case and
in the lattice situation (i.e. with account of spin dynamics). In Sect.3,
results of numerical calculations are presented. In Sect.4 we discuss the
physical consequences. Details of derivation of the scaling equations are
presented in Appendices.

\section{Scaling equations}

To describe a Kondo lattice, we use the degenerate-band (multichannel)
periodical $s-f$ exchange model
\begin{equation}
H=\sum_{\mathbf{k}m\sigma }t_{\mathbf{k}}c_{\mathbf{k}m\sigma }^{\dagger }c_{%
\mathbf{k}m\sigma }-I\sum_{im\sigma \sigma ^{\prime }}\mathbf{S}_{i}%
\mbox {\boldmath
$\sigma $}_{\sigma \sigma ^{\prime }}c_{im\sigma }^{\dagger }c_{im\sigma
^{\prime }}^{{}}+H_{f}  \label{1}
\end{equation}%
where $t_{\mathbf{k}}$ is the band energy, $\mathbf{S}_{i}$ are spin-1/2
operators, $I$ is the $s-f$ exchange parameter, $\sigma $ are the Pauli
matrices, $m=1...M$ is the channel index. For the sake of convenient
constructing perturbation theory, we explicitly include the Heisenberg $f-f$
exchange interaction $H_{f}$ in the Hamiltonian, although in fact this
interaction is usually the indirect RKKY coupling.

In the more general $SU(N)\otimes SU(M)$ model we have $\sigma =1...N$ and
the Hamiltonian can be written as \cite{Cox}

\begin{equation}
H=\sum_{\mathbf{k}m\sigma }t_{\mathbf{k}}c_{\mathbf{k}m\sigma }^{\dagger }c_{%
\mathbf{k}m\sigma }^{{}}-I\sum_{im\sigma \sigma ^{\prime }}|i\sigma ^{\prime
}\rangle \langle i\sigma |c_{im\sigma }^{\dagger }c_{im\sigma ^{\prime
}}^{{}}+H_{f}  \label{1llll}
\end{equation}%
A somewhat more realistic model including angular momenta is discussed in
Ref.\cite{IK}; generalization to arbitrary spin is also possible (see,
e.g.,. Ref.\cite{507nfl})

Similar to Ref.\cite{IK} we use the ``poor man
scaling'' approach \cite{And}. In this method one considers
the dependence of effective (renormalized) model parameters on the cutoff
parameter $C<0$ which occurs at picking out the Kondo singular terms and
approaching the Fermi level.

To describe the renormalization process we introduce the dimensionless
coupling constants
\begin{equation}
g_{ef}(C)=-N\rho I_{ef}(C),\quad g=-N\rho I
\end{equation}%
where $\rho$ is the bare electron density of states per channel at the Fermi level.  In the one-impurity case the scaling behaviour is governed by the beta
function%
\begin{equation}
\beta (g)=-g^{2}+(M/N)g^{3}+...
\end{equation}%
At $M>N~$the fixed point $g^{\ast }=N/M$ (zero of $\beta (g)$) lies in a
weak coupling region which makes possible successful application of
perturbation and renormalization group approaches.

The scaling equation reads
\begin{equation}
\int_{g}^{g_{ef}(C)}\frac{dg}{\beta (g)}=\int_{D}^{C}\frac{dC}{C}=\ln
\left\vert \frac{C}{D}\right\vert
\end{equation}%
where the cutoff energy $D$ is defined by $g_{ef}(-D)=g$. Solving this
equation yields%
\begin{equation}
\frac{g^{\ast }-g_{ef}(C)}{g^{\ast }-g}=g^{\ast }\left\vert \frac{C}{T_{K}}%
\right\vert ^{\Delta }\exp \left( -\frac{g^{\ast }}{g_{ef}(C)}\right)
\label{pow}
\end{equation}%
with $\Delta =N/M$ and the Kondo temperature%
\begin{equation}
T_{K}=Dg^{M/N}\exp (-1/g)  \label{TK}.
\end{equation}%
It should be noted that we have no divergence of $g_{ef}(\xi )$, and the
power-law critical behavior in (\ref{pow}) takes place in a wide region,
including $|C|>T_{K}$ \cite{Gan}.

Generally, the critical exponents are defined by the slope $\Delta =\beta
^{\prime }(g).$ Taking into account higher orders in $1/M$ one has
\begin{equation}
\Delta =\frac{N}{M}\left( 1-\frac{N}{M}\right) \simeq \frac{N}{M+N},
\label{Delts}
\end{equation}%
the latter value being in agreement with the exact results of Bethe ansatz
and conformal field theory (see Ref. \cite{Cox}). The corresponding value of
$g^{\ast }$ for $N=2$ reads \cite{Gan}%
\begin{equation}
g^{\ast }=\frac{2}{M}\left( 1-\frac{2\ln 2}{M}\right)
\end{equation}%
which differs weakly from $\Delta .$

Using the results of Appendices A, B we can write down the system of scaling
equations for paramagnetic (PM), ferromagnetic (FM) and antiferromagnetic
(AFM) phases in the lattice case. Similar to Ref.\cite{IK}, but taking into
account next-leading contributions,
we find the equation for $I_{ef}$ by picking
up in the sums in the corresponding self-energies the
contribution of intermediate electron states near the Fermi level with $C<t_{%
\mathbf{k}}<C+\delta C$. We derive%
\begin{equation}
\delta I_{ef}(C)=\rho I^{2}N(1+M\rho I)\eta (-\frac{\overline{\omega }}{C}%
)\delta C/C  \label{sef}
\end{equation}%
where $\overline{\omega }$ is a characteristic spin-fluctuation energy, $%
\eta (x)$ is the scaling function satisfying the condition $\eta (0)=1,$
which guarantees the correct one-impurity limit, see Appendix C.
The third-order term, proportional to $M$, comes from corrections which contain summation over the orbital index $m$ (in the diagram approach, they correspond to diagrams containing a closed electron loop).

The leading renormalization of spin-fluctuation frequencies is already of order of $M$:
\begin{equation}
\delta \overline{\omega }_{ef}(C)/\overline{\omega }=a\delta \overline{S}%
_{ef}(C)/S=aMN\rho ^{2}I^{2}\eta (-\frac{\overline{\omega }}{C})\delta C/C
\label{wef}
\end{equation}%
where the parameters $a$ for a concrete lattice and magnetic structure are expressed in terms of averages over the Fermi surface (see Refs.\cite{IK,731} and Appendices A, B).
 It turns out that, owing to the structure of perturbation theory for magnetic characteristics, the   $M^2$ corrections do not occur in the third order in $I$, so that Eq.\ref{wef} is sufficient.

Replacing in the right-hand parts of (\ref{sef}) and (\ref{wef}) $%
g\rightarrow g_{ef}(C)$, $\overline{\omega }\rightarrow \overline{\omega }%
_{ef}(C)$ we obtain the system of scaling equation
\begin{equation}
[1-\gamma g_{ef}(C)]^{-1}\partial g_{ef}(C)/\partial C =-\Lambda
\label{gl}
\end{equation}%
\begin{equation}
a\partial \ln \overline{S}_{ef}(C)/\partial C =\partial \ln \overline{%
\omega }_{ef}(C)/\partial C=a\gamma \Lambda  \label{wl}
\end{equation}%
with $\gamma =M/N,$
\begin{equation}
\Lambda =\Lambda (C,\overline{\omega }_{ef}(C))=[g_{ef}^{2}(C)/C]\eta (-%
\overline{\omega }_{ef}(C)/C).
\end{equation}%
Writing down the first integral of the system (\ref{gl}), (\ref{wl}) yields

\begin{equation*}
\ln |1-\gamma g_{ef}(C)|-(1/a)\ln \overline{\omega }_{ef}(C)=\operatorname{const}
\end{equation*}%
\begin{equation}
\frac{\overline{S}_{ef}(C)}{S} =\left( \frac{\overline{\omega }_{ef}(C)}{%
\overline{\omega }}\right) ^{1/a}=\frac{1-\gamma g_{ef}(C)}{1-\gamma g}=%
\frac{g^{\ast }-g_{ef}(C)}{g^{\ast }-g},  \label{w+g} \\
\end{equation}%
\begin{equation}
(M/N)g_{ef}(C) =1-[1-(M/N)g]\left( \frac{\overline{\omega }_{ef}(C)}{%
\overline{\omega }}\right) ^{1/a}
\end{equation}%
Thus we have a soft-mode situation at approaching the fixed point.

Provided that $\overline{\omega }_{ef}(C)$ is weakly renormalized (e.g., $%
a\ll 1$ at smalll $k_{F}$) we obtain

\begin{equation}
\frac{g^{\ast }-g_{ef}(C)}{g^{\ast }-g} =g^{\ast }\exp \left( -\frac{%
g^{\ast }}{g_{ef}(C)}\right) \left\vert \frac{D\exp [G(C)]}{T_{K}}%
\right\vert ^{\Delta }
\end{equation}%
\begin{equation}
G(C) =-\int_{-D}^{C}\frac{dC^{\prime }}{C^{\prime }}\eta \left( -\frac{%
\overline{\omega }}{C^{\prime }}\right) ,
\end{equation}%
cf. the treatment of the large-$N$ limit \cite{IK}. In particular, in the
paramagnetic state%
\begin{equation*}
G^{PM}(C)=\frac{1}{2}\ln ((C^{2}+\overline{\omega }^{2})/D^{2})+\frac{C}{%
\overline{\omega }}\arctan (\frac{\overline{\omega }}{C})-1
\end{equation*}%
so that $C\rightarrow \sqrt{C^{2}+\overline{\omega }^{2}}$ in the Kondo
divergences in comparison with (\ref{pow}), cf. discussion in Refs.\cite%
{IKFTT,IKFTT1}.

However, in the general case the scaling behavior is much more rich and
interesting. Introducing the function
\begin{equation}
\chi (\xi )=\ln \frac{\overline{\omega }}{\overline{\omega }_{ef}(\xi )}%
=a\ln \frac{\overline{S}_{ef}(C)}{S}
\end{equation}%
the scaling equation takes the form

\begin{equation}
\frac{\partial \chi }{\partial \xi }=\frac{a}{\gamma }\left[ 1-(1-\gamma
g)\exp (-\chi /a)\right] ^{2}\Psi (\lambda +\chi -\xi )  \label{linf}
\end{equation}%
where
\begin{equation*}
\Psi (\xi )=\eta (e^{-\xi }),\quad \xi =\ln |D/C|,\quad \lambda =\ln (D/\overline{%
\omega })\gg 1
\end{equation*}%
\qquad \qquad

In Ref.\cite{IK}, an approximation was proposed for magnetically ordered
cases, which takes into account not only the magnon pole, but also
incoherent contribution, namely

\begin{multline}
\Lambda =[g_{ef}^{2}(C)/C] \\
\times [Z\eta _{coh}(-\overline{\omega }_{ef}(C)/C)+(1-Z)\eta _{incoh}(-\overline{\omega }_{ef}(C)/C)]  
\label{l1}
\end{multline}%
where $\eta _{coh}$ corresponds to the magnetic phase, and the function $\eta
_{incoh}$ is unknown; for estimations we may put $\eta _{incoh}=\eta ^{PM}.~$%
The quantity $Z=Z(-\overline{\omega }_{ef}(C)/C)$ is the residue at the
magnon pole, which is given by

\begin{equation}
\frac{1}{Z(\xi )}=1+\ln \frac{S}{\overline{S}(\xi )}  \label{1/ZS}
\end{equation}%
Then we have instead of (\ref{linf})%
\begin{multline}
\frac{\partial \chi }{\partial \xi }=\frac{a}{\gamma }\left[ 1-(1-\gamma
g)\exp (-\chi /a)\right] ^{2} \\ \times [Z\Psi _{coh}(\lambda +\chi -\xi )+(1-Z)\Psi
_{incoh}(\lambda +\chi -\xi )]  \label{wzll}
\end{multline}%
with $Z=1/(1+\chi /a).$

\section{Scaling behavior}

Our scaling equations are written in terms of $\gamma $ rather than $M$ and $%
N$ separately. Therefore, to establish properly the correspondence with the
one-impurity case (\ref{Delts}), we may put $\gamma =M/N+1=1/\Delta .$ This
yields, at least for $M>2$, correct critical exponents for magnetic
susceptibility, specific heat and resistivity.

The important case $M=2$ is more difficult from the theoretical point of
view, see \cite{Cox,Col,Parcollet}. However, a fixed point is still present
for $M=2,$ the resistivity being satisfactorily described by simple scaling
approach \cite{Cox}.

In numerical calculations, we put $M=3,$ which may be relevant for Ce$^{3+}$
ion \cite{Cox}. Then for $M=3,~N=2$ we have $\gamma =5/2.$

Since $\Psi (\xi >1)\simeq 1,$ in the PM phase $\chi (\xi )$ increases
according to (\ref{pow}), (\ref{w+g}). Provided that $g$ is not too small,
at large $\xi $ we can put for rough estimations $g_{ef}^{{}}(\xi )\simeq
g^{\ast }=1/\gamma $ to obtain%
\begin{equation}
\chi (\xi )\simeq a\gamma g_{ef}^{2}(\xi )\xi -a/\gamma g\simeq (a/\gamma
)(\xi -1/g).  \label{chi}
\end{equation}%
Thus a power-law behavior occurs
\begin{eqnarray}
\overline{\omega }_{ef}(C) &\simeq &\overline{\omega }(|C|/T_{K})^{\beta
},\quad \beta =a/\gamma =a\Delta  \nonumber \\
\overline{S}_{ef}(C) &\simeq &(|C|/T_{K})^{\Delta },~  \label{pm}
\end{eqnarray}%
which corresponds to the standard one-impurity NFL behavior (see below the
discussion of physical properties). Note that the scale of $T_{K}$ occurs
here, unlike the lowest-order scaling in the large-$M$ limit \cite{IK}. The
dependence (\ref{chi}) takes place up to the point
\begin{equation}
\xi _{1}\simeq (\lambda -\beta /g)/(1-\beta ).  \label{bound}
\end{equation}%
For $\xi >\xi _{1},$~$\chi (\xi )\simeq \chi (\xi _{1})\simeq \lambda \beta
/(1-\beta )$ is practically constant since $\Psi (\lambda +\chi -\xi )$
becomes small, and $g_{ef}(\xi )$ increases slowly tending at $\xi
\rightarrow \infty $ to an asymptotic value which is, however, smaller than
the one-impuirity $g^{\ast }$ since $\chi (\xi )$ remains finite.

Note that lowest-order (one-loop) scaling for finite $M$ yields the NFL behavior in a very
narrow interval of bare coupling constant $g$ only, since with increasing $g$
we come rapidly to strong-coupling regime where $g_{ef}(\xi >\lambda
)\rightarrow \infty .$ Unlike the lowest-order scaling, such a
critical $g$ value does not occur in the present calculation for the paramagnetic case: $g_{ef}(\xi )$ remains finite for any $g$.

\begin{figure}[tbp]
\includegraphics[width=3.3in, angle=0,clip]{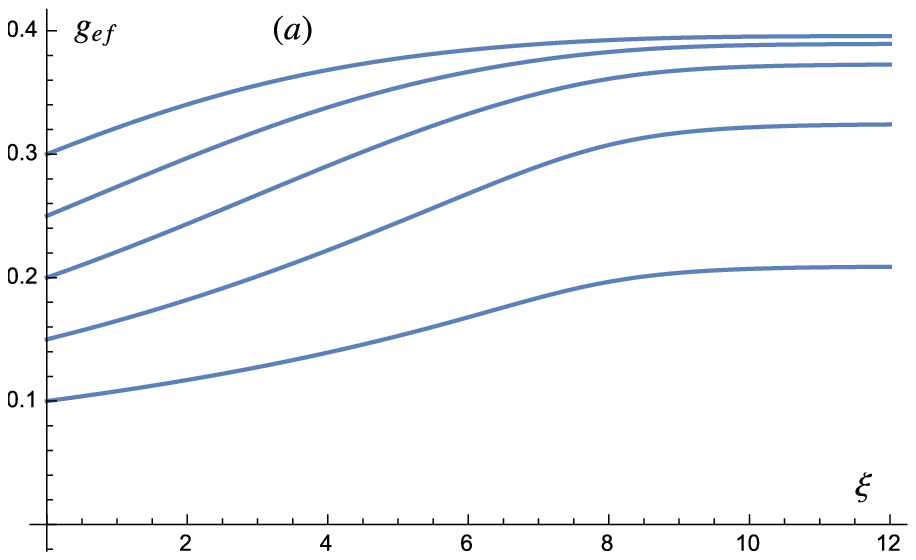}
\includegraphics[width=3.3in, angle=0,clip]{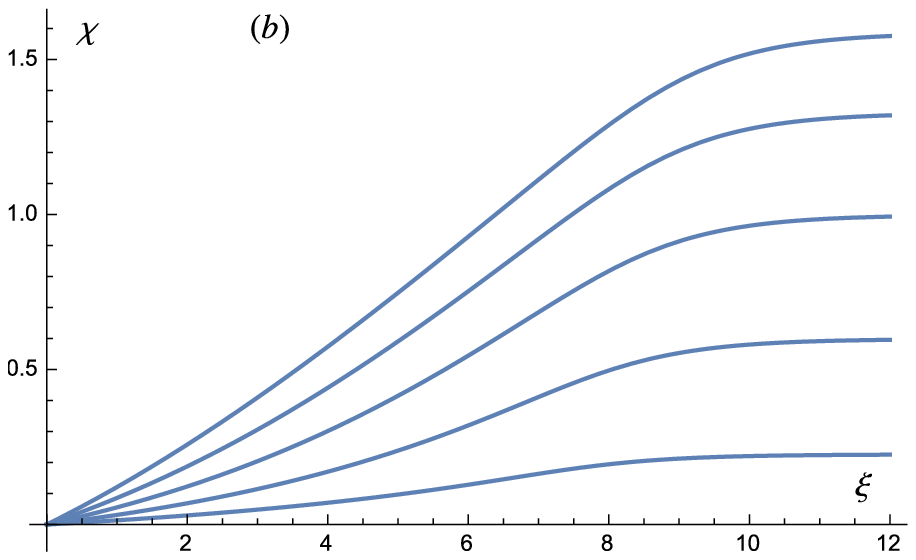}
\caption{
The scaling trajectories for a 3D paramagnet, $g_{ef}(\xi )$ (a) and
$\chi (\xi )$ (b). The parameter values are $\lambda =7$, $a=0.5,$ $%
g=0.1,0.15,0.2,0.25,0.3$ (for the curves from below to above)
}
\label{fig:1}
\end{figure}

The dependences $g_{ef}(\xi )$ and $\chi (\xi )$ in the paramagnetic phase
are shown in Fig.1 for the 3D case (the results for the 2D case differ here
very weakly). The behavior $g_{ef}(\xi )$ between $\xi _{1}$ and $\xi _{2}$
may be described as nearly linear, but is somewhat smeared since $\Psi (\xi
) $ differs considerably from the asymptotic values 0 and 1 in a rather
large interval of $\xi .$ 
Remember that Fig.1a demonstrates also the behavior of magnetic moment according to Eq.(\ref{w+g}).

In magnetically ordered phases, the behavior for for $\xi <\xi _{1}$ is
similar, but the situation for $\xi >\xi _{1}$ changes since the Van Hove singularity
of $\Psi _{coh}(\xi )$ at $\xi =0$ plays an important role. Instead of
decreasing, $\Psi (\lambda +\chi -\xi )$ starts to increase at approaching $%
\xi _{1}$. 
At sufficiently large $g$, provided that
\begin{equation}
a\gamma g_{ef}^{2}(\xi \simeq \xi _{1})\Psi _{coh}^{\max }\simeq a\gamma
g^{\ast 2}\Psi _{coh}^{\max }>1,
\end{equation}%
at $\xi >\xi _{1}$ the argument of the function $\Psi _{coh}$ in (\ref{linf}%
) becomes almost constant (fixed), and we obtain
\begin{equation}
\chi (\xi )\simeq \xi -\lambda ,\;\overline{\omega }_{ef}(C)\simeq |C|.
\label{lin}
\end{equation}%
Thus, instead of divergence of $\chi (\xi )$ in the one-channel model \cite%
{IK}, we have a linear NFL behavior since $g_{ef}(\xi )$ remains finite.
Such a behavior has a critical nature and corresponds to $g=g_{c}$ in the
one-channel Kondo model.

Unlike the PM case, a sharp crossover occurs here with changing $g$ since we
do not reach the regime (\ref{lin}) at small $g<g_{c}$. The value of $g_{c}$
is determined by the value of $\delta ,$ i.e. the magnon damping, see (\ref%
{delta}). One can see that the influence of the singularity is considerably
stronger and conditions of the NFL behavior are more favorable in the 2D
rather than 3D case, and in the AFM rather than FM case. Above the critical
value $g_{c}$, the picture of scaling trajectories (in particular, the size
of NFL behavior region ) does not practically depend on $g$.

In the case of equation (\ref{linf}), the linear behavior takes place up to $%
\xi =\infty $. On the other hand, when taking into account the incoherent
contribution the increase of $\chi $ stops at $a\gamma g^{\ast 2}\Psi
_{coh}^{\max }=1/Z=1+\chi /a,$ i.e., at
\begin{equation}
\xi _{2}=\lambda +\chi _{\max }=\lambda +a(a\gamma g^{\ast 2}\Psi
_{coh}^{\max }-1).
\end{equation}%

\begin{figure}[tbp]
\includegraphics[width=3.3in, angle=0]{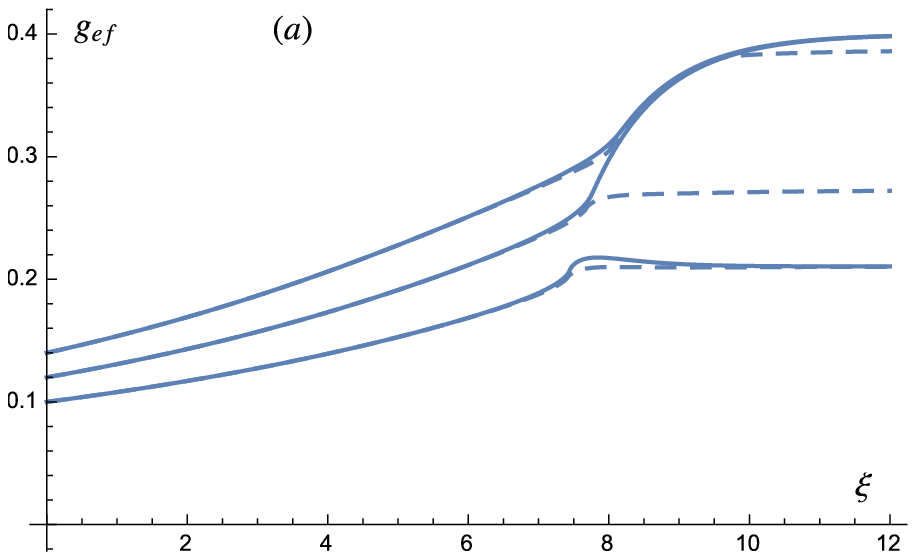}
\includegraphics[width=3.3in, angle=0]{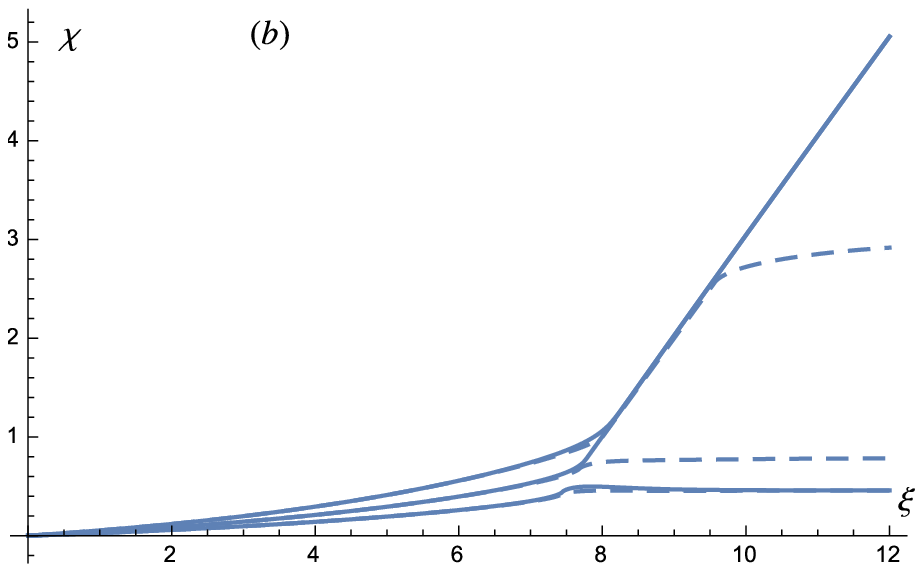}
\caption{
The scaling trajectories for a 3D antiferromagnet $g_{ef}(\xi )$ (a)
and $\chi (\xi )$ (b). The solid lines correspond to the approximation (\ref%
{linf}), and the dashed lines to account of incoherent contribution, Eq. (%
\ref{wzll}). The parameter values are $\lambda =7$, $\delta =2~10^{-4},$ $%
a=1 $, $g=0.1,0.12,0.14$ (from below to above). Note a shallow maximum of $%
g_{ef}(\xi )$ for small $g$, which is due to the sign change in $\eta (x)$
}
\label{fig:2}
\end{figure}

\begin{figure}[tbp]
\includegraphics[width=3.3in, angle=0]{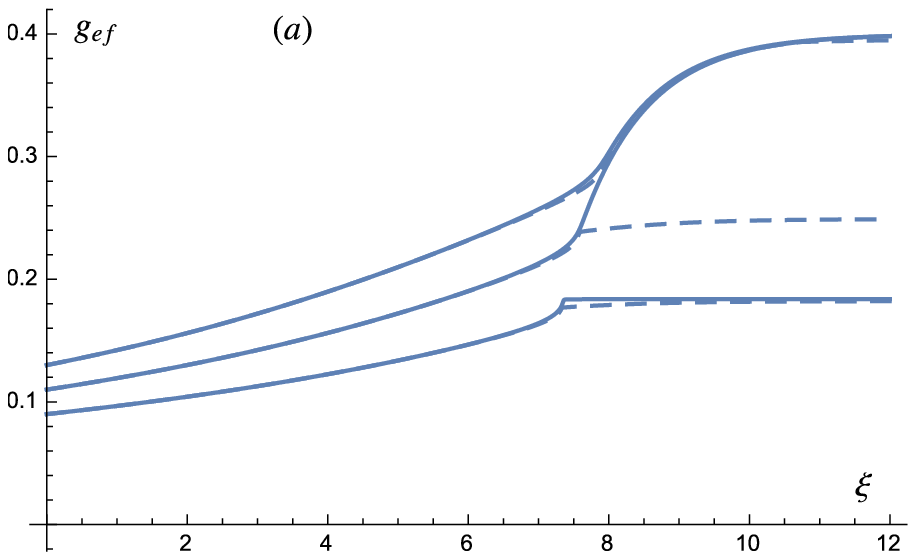}
\includegraphics[width=3.3in, angle=0]{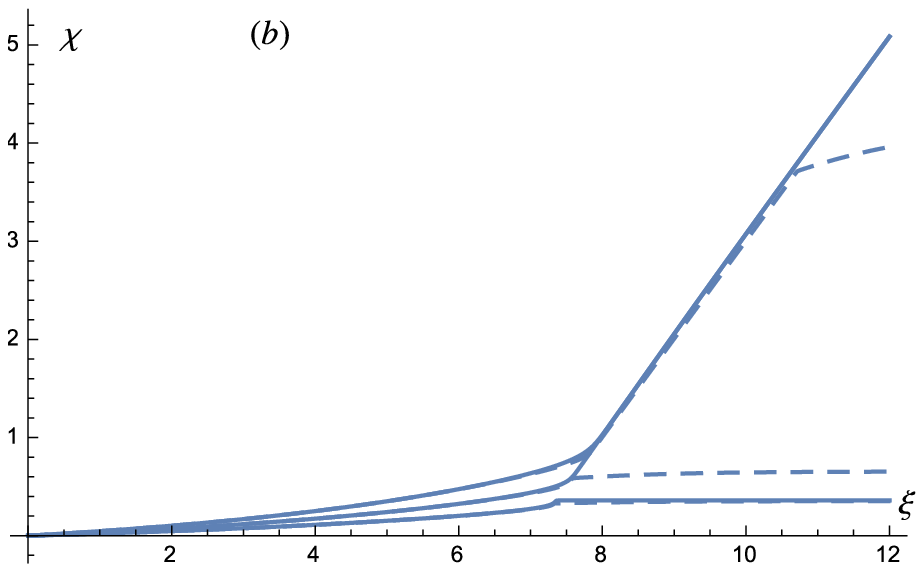}
\caption{
The scaling trajectories for a 2D antiferromagnet $g_{ef}(\xi )$ (a)
and $\chi (\xi )$ (b). The parameter values are $\lambda =7$, $\delta
=4~10^{-3},$ $a=1$, $g=0.09,0.11,0.13$ (from below to above). The value $%
g=0.11$ is intermediate: $g>g_{c}$ for the approximation (\ref{linf}) and $%
g<g_{c}$ for the approximation (\ref{wzll})
}
\label{fig:3}
\end{figure}

The dependences $\chi (\xi )$ for a 3D and 2D antiferromagnet are shown in
Figs.2-3. In the presence of the incoherent contribution, the region, where
the linear dependence (\ref{lin}) holds, is sensitive to the value of $%
\delta $ and is not  wide, especially in the 3D case; the width does not
increase with further increasing $g$. However, a more exact consideration of
spin dynamics may change considerably the results. Probably, using the spin
diffusion approximation underestimates the coherence and the picture should
be somewhat intermediate between solid and dashed lines.

\section{The non-Fermi-liquid behavior in physical properties}

Now we discuss the NFL behavior of physical properties for the most
important case $N=2$. The temperature dependences of magnetic moment and
magnetic susceptibility In the PM case are obtained directly from the above
results by the replacement $|C|\rightarrow T$,%
\begin{equation}
S_{ef}^{{}}(T)\varpropto (T/T_{K})^{\Delta },~\chi _{m}\varpropto
S_{ef}^{2}(T)/T\varpropto (T/T_{K})^{2\Delta }/T.
\end{equation}%
However, unlike the one-impurity case, such dependences are somewhat smeared
and take place only up to temperatures determined by (\ref{bound}). A
similar dependence is obtained for specific heat \cite{Gan}.

As discussed above, the logarithmic factor in $\chi _{m}$ for $M=2$ ($\Delta
=1/2$) is not described by our approach; an accurate treatment is obtained
by more sophisticated methods, e.g., Bethe ansatz and conformal field
theory.

In the spin-wave region for an AFM structure with the wavevector $\mathbf{Q}$
we write down in terms of a retarded Green's function
\begin{equation}
\chi _{m}=\lim_{q\rightarrow 0}\langle \langle S_{\mathbf{q}}^{x}|S_{-%
\mathbf{q}}^{x}\rangle \rangle _{\omega =0}\propto \overline{S}/\overline{%
\omega }.
\end{equation}%
On replacing $\overline{\omega }\rightarrow \overline{\omega }(C),$ $%
\overline{S}\rightarrow \overline{S}_{ef}(C)$ with $|C|\sim T$ in spirit of
scaling arguments we obtain
\begin{equation}
\chi_m (T)\propto T^{-\zeta },~\zeta =\left\{
\begin{array}{c}
(a-1)/a \\
\Delta (a-1)/a%
\end{array}%
\right.
\end{equation}%
for the regimes corresponding to (\ref{pm}) and (\ref{lin}), respectively.
Note that the spin-wave description of the electron-magnon interaction can
be adequate not only in the AFM phase, but also for systems with a strong
short-range magnetic order, including 2D and frustrated 3D systems at finite
temperatures.

According to (\ref{alph}), the non-universal exponent $\zeta $ is determined
by details of magnetic structure and can be both positive and negative. For
a qualitative discussion, we can use Figs.2-3 and treat the difference $a-1$
as a perturbation.

Following to Ref. \cite{731}, the temperature dependence of electronic
specific heat in magnetic phases can be estimated from the second-order
perturbation theory, $C_{el}(T)/T\propto 1/z(T)$ where $z(T)$ is the residue
of the electron Green's function at the distance $T$ from the Fermi level
(cf. Ref.\cite{IKFTT}). Then we have in the AFM case%
\begin{equation}
C_{el}(T)/T\propto g_{ef}^{2}(T)\overline{S}_{ef}(T)/\overline{\omega }%
_{ef}(T)\propto \chi _{m}(T).
\end{equation}%
The dependence $C_{el}(T)/T\propto \chi _{m}(T)$ was obtained experimentally
for a number of NFL systems \cite{Tsv,Cox}.

In the paramagnetic case the temperature correction to magnetic resistivity
can be calculated from (\ref{pow}) as \cite{Gan}%
\begin{equation}
\delta R_{m}(T)\propto g_{ef}(T)-g^{\ast }\propto -(T/T_{K})^{\Delta }.
\end{equation}%
The $T^{1/2}~$dependence (which corresponds to $M=2$) is indeed observed in
a number of $f$-systems \cite{Cox}.

In the regime (\ref{lin}), the contribution to resistivity owing to
scattering by spin fluctuations in AFM phase is given by \cite{731}
\begin{equation}
\delta R(T)\propto T^{2}g_{ef}^{2}(T)\overline{S}_{ef}(T)/\overline{\omega
}_{ef}(T)\propto T^{2}C_{el}(T)/T\propto T^{2-\zeta }.  \label{taul}
\end{equation}%
For electron-electron scattering one has another temperature dependence

\begin{equation}
\delta R(T)\propto T^{2}\left[ C_{el}(T)/T\right] ^{2}\propto T^{2-2\zeta }.
\label{tau2}
\end{equation}


\section{Conclusions}

To conclude, we have treated various mechanisms of the NFL behavior in the
multichannel Kondo lattice. In comparison with one-impurity model, the
lattice version provides a more rich picture. The NFL phenomenon seems to
have a complicated nature being influenced by both single-site Kondo effect
and spin dynamics. The corresponding  dependences of physical
properties can be different in different temperature intervals. Moreover, various
scattering mechanisms can give different temperature dependences.

The most important result is occurence of an intermediate-coupling fixed point, which means formation of  reduced magnetic moment or even its vanishing in the NFL regime, the dependence on the bare coupling parameter being weak. 
The details of scaling behavior are determined by the magnetic structure
(parameter $a$) and the scaling function $\eta (x)$, its singularities being essential. Peculiarities of electron and magnon spectrum can also play a role, similar to consideration in Refs.\cite%
{731,Infl}.


An important problem is stability of the fixed point: lifting of the degeneracy of electron subbands with different $m$ in the
Hamiltonian (\ref{1}) should result in a change of scaling behavior, so that anomalous temperature dependences may take place in a restricted region.
Possible applications of two-channel model to rare-earth and actinide
systems, including corresponding difficulties of interpretation, are
discussed in Ref. \cite{Cox}. For uranium systems, realization of this model
is possible due to time-reversal symmetry of subbands.


The model used describes naturally formation of a magnetic state with small
moment value. Besides that, our consideration provides an example of
essential renormalization of the coupling constant according to (\ref{w+g}).
This may be of interest for the general theory of metallic magnetism (in
particular, for weak itinerant ferro- and antiferromagnets): the magnetic
state is determined by the renormalization process rather than by bare
Stoner-like criterion (cf. discussion in Ref.\cite{IKFTT1}).

The author is grateful to Prof. M. I. Katsnelson for useful discussions.
This work was supported in part by the Division of Physical Sciences and Ural Branch of Russian Academy of Sciences (project no. 15-8-2-9).

\section*{Appendix A. Renormalization in the paramagnetic phase}

The Kondo-lattice problem in the paramagnetic state describes the process of
screening of localized magnetic moments. The correction to the effective
magnetic moment  is obtained from the static magnetic susceptibility
\cite{IKFTT,IK}%
\begin{eqnarray}
\chi_m &=&\overline{S}_{ef}^{2}/3T,\quad \overline{S}_{ef}^{2}=S(S+1)[1-L]
\label{Sef} \\
L &=&4MNI^{2}\int_{-\infty }^{\infty }d\omega \sum_{\mathbf{kq}}\mathcal{J}_{%
\mathbf{q}}(\omega )\frac{n_{\mathbf{k}}(1-n_{\mathbf{k-q}})}{(t_{\mathbf{k}%
}-t_{\mathbf{k-q}}-\omega )^{2}}  \nonumber
\end{eqnarray}%
with $\mathcal{J}_{\mathbf{q}}(\omega )$ the spectral density of the spin
Green's function for the Hamiltonian $H_{f},$ which is normalized to unity.
We use the simple spin diffusion approximation%
\begin{equation}
\mathcal{J}_{\mathbf{q}}(\omega )=\frac{1}{\pi }\frac{\mathcal{D}q^{2}}{%
\omega ^{2}+(\mathcal{D}q^{2})^{2}}  \label{JD}
\end{equation}%
($\mathcal{D}$ is the spin diffusion constant), which corresponds to
dissipative spin dynamics.

The spin-fluctuation frequency in the paramagnetic phase is determined from
the second moment of the spin Green's function with the result \cite{IKFTT,IK}
\begin{equation}
\delta \omega _{\mathbf{q}}^{2}/\omega _{\mathbf{q}}^{2}=(1-\alpha _{\mathbf{%
q}})\delta \overline{S}_{ef}^{2}/\overline{S}_{ef}^{2}=-(1-\alpha _{%
\mathbf{q}})L  \label{dwpm}
\end{equation}%
where $\alpha _{\mathbf{q}}$ is expressed in terms of exchange integrals%
\begin{equation}
\alpha _{\mathbf{q}}=\sum_{\mathbf{R}}J_{\mathbf{R}}^{2}\left( \frac{\sin
k_{F}R}{k_{F}R}\right) ^{2}[1-\cos \mathbf{qR]/}\sum_{\mathbf{R}}J_{\mathbf{R%
}}^{2}[1-\cos \mathbf{qR]}  \label{Alpq}
\end{equation}%
In the approximation of nearest neighbors at the distance $d,$
\begin{equation}
\alpha _{\mathbf{q}}=\alpha =\langle e^{i\mathbf{kR}}\rangle _{t_{\mathbf{k}%
}=E_{F}}=\left( \frac{\sin k_{F}d}{k_{F}d}\right) ^{2},  \label{ALp}
\end{equation}%
so that we may use a single renormalization parameter. 

It should be stressed
that we do not need here to search for higher order corrections to magnetic properties (leading corrections are already proportional to $M$).

To construct a self-consistent theory of Kondo lattices we have
to find the
renormalization of the effective $s-f$ exchange parameter. To this end, we
calculate
the Kondo correction to the electron self-energy with account of spin dynamics. We 
use the method of irreducible Green's functions (see Ref.\cite{329} 
and the review paper \cite{329a}) which enables one to construct a consistent perturbation expansion in a small parameter. We write down
\begin{equation}
\Sigma _{\mathbf{k}}^{{}}(E)=\langle \!\langle \lbrack c_{\mathbf{k\sigma }%
}^{{}},H_{int}]|[H_{int},c_{\mathbf{k\sigma }}^{\dagger }]\rangle \rangle
_{E}^{irr}.
\end{equation}
where $H_{int}$ is the $s-f$ interaction term. In the second order in $I$ we have
\begin{equation}
\Sigma _{\mathbf{k}}^{(2)}(E)=I^{2}PR(E),~R(E)=\sum_{\mathbf{q}}\frac{1}{%
E-t_{\mathbf{k-q}}}.  \label{sig2}
\end{equation}%
The next-order singular contributions read%
\begin{multline}
\Sigma _{\mathbf{k}}^{(3)}(E)=-I^{3}PN\int_{-\infty }^{\infty }d\omega \sum_{%
\mathbf{q,p}}\mathcal{J}_{\mathbf{q}}(\omega )\frac{n_{\mathbf{k-q}}}{E-t_{%
\mathbf{k-q}}-\omega } \\ 
\times \left( \frac{1}{E-t_{\mathbf{k-p}}}-\frac{1}{t_{%
\mathbf{k-q}}-t_{\mathbf{k-p}}}\right)  \label{sig3}
\end{multline}%
\begin{multline}
\Sigma _{\mathbf{k}}^{(4)}(E)=-I^{4}PMN\sum_{\mathbf{q,p}}\int_{-\infty
}^{\infty }d\omega d\omega ^{\prime }\mathcal{J}_{\mathbf{q}}(\omega )%
\mathcal{J}_{\mathbf{q-p}}(\omega ^{\prime }) \\
\times \frac{1}{(E-t_{\mathbf{k-q}%
}-\omega )^{2}}\frac{n_{\mathbf{k}^{\prime }}(1-n_{\mathbf{k}^{\prime }%
\mathbf{-p}})}{E-t_{\mathbf{k-q}}+t_{\mathbf{k}^{\prime }}-t_{\mathbf{k}%
^{\prime }\mathbf{-p}}-\omega ^{\prime }}  \label{sig4}
\end{multline}%
where $P=1-1/N^{2},$ $n_{\mathbf{k}}=n(t_{\mathbf{k}})$ is the Fermi function. 
When neglecting
spin dynamics Eqs. (\ref{sig2})-(\ref{sig4}) agree with the one-impurity
results \cite{Gan}. The Kondo renormalization of the $s-f$ parameter $%
I\rightarrow I_{ef}=I+\delta I_{ef}$ is determined by ``incorporating'' $\operatorname{Im}\Sigma _{\mathbf{k}}^{(3)}(E)$ into  $\operatorname{Im}\Sigma _{\mathbf{k}}^{(2)}(E).$ The  imaginary parts required are simplified:%
\begin{multline}
-\operatorname{Im}\Sigma _{\mathbf{k}}^{{}}(E) =\pi I^{2}\rho P \\
\times \left( 1-IN\int_{-\infty }^{\infty }d\omega \sum_{\mathbf{q,p}}\mathcal{J}_{\mathbf{%
q}}(\omega )\frac{n_{\mathbf{k-q}}}{E-t_{\mathbf{k-q}}-\omega }\right. \\
-MNI^{2}\int_{-\infty }^{\infty }\int_{-\infty }^{\infty }d\omega
d\omega ^{\prime } \\
\left.\times \sum_{\mathbf{q,p}}\mathcal{J}_{\mathbf{k-k}^{\prime
}}(\omega )\mathcal{J}_{\mathbf{k-k}''}(\omega ^{\prime })\frac{n_{\mathbf{k}%
^{\prime }}(1-n_{\mathbf{k}''})}{(t_{\mathbf{k}^{\prime }}-t_{\mathbf{k}%
''}-\omega +\omega ^{\prime })^{2}}\right).
\end{multline}%
Note that the structure of $\operatorname{Im}\Sigma _{\mathbf{k}}^{(4)}(E)$ is
similar to that for magnetic susceptibility and magnetic moment (\ref{Sef}).
Averaging over $t_{\mathbf{k}}=t_{\mathbf{k}^{\prime }}=t_{\mathbf{k}%
"}=E_{F}=0$ we obtain to leading accuracy the result (\ref{sef}).

\section*{Appendix B. Renormalization in magnetically ordered phases}

Now we investigate the renormalization of the $s-f$ interaction in FM and
AFM phases. For simplicity we treat only the $s-f$ model with $N=2$ (a more
general case is discussed in Ref.\cite{IK}).

For a ferromagnet the electron spectrum possesses the spin splitting, $E_{%
\mathbf{k}\sigma }=t_{\mathbf{k}}-\sigma MI\overline{S}.$ The second-order
correction to $I_{ef}$ is determined by the corresponding electron
self-energies:
\begin{equation}
\delta I_{ef}=[\Sigma _{\mathbf{k\downarrow }}^{FM}(E)-\Sigma _{\mathbf{%
k\uparrow }}^{FM}(E)]/(2\overline{S})  \label{Ieffm}
\end{equation}%
which are defined by%
\begin{equation*}
\sum\limits_{mm^{\prime }}\langle \langle c_{\mathbf{k}m\sigma }^{{}}|c_{%
\mathbf{k}m^{\prime }\sigma }^{\dagger }\rangle \rangle _{E}=M/[E-t_{\mathbf{k%
}}-\sigma I\overline{S}-\Sigma _{\mathbf{k\sigma }}^{{}}(E)]
\end{equation*}%
As described in Ref.\cite{329}, using equation-of-motion method, we write down the
self-energy in terms of the irreducible Green's function%
\begin{multline}
\Sigma _{\mathbf{k\sigma }}^{{}}(E)=I^{2}\sum_{m^{\prime }\mathbf{q}}\langle
\!\langle S_{\mathbf{q}}^{-\mathbf{\sigma }}c_{\mathbf{k-q}m\mathbf{-\sigma }%
}^{{}}+\sigma \delta S_{\mathbf{q}}^{z}c_{\mathbf{k-q}m\mathbf{\sigma }%
}^{{}}|\\
c_{\mathbf{k+p}m^{\prime }\mathbf{-\sigma }}^{\dagger }S_{\mathbf{p}%
}^{\mathbf{\sigma }}+\sigma c_{\mathbf{k+p}m^{\prime }\mathbf{\sigma }%
}^{\dagger }\delta S_{\mathbf{p}}^{z}\rangle \rangle _{E}^{irr}  \label{sirr}
\end{multline}%
with $\delta A=A-\langle A\rangle .$ Writing down the equations of motion
for the Green's function (\ref{sirr}) we derive with account of singular
terms%
\begin{eqnarray}
\Sigma _{\mathbf{k\uparrow }}^{{}}(E) &=&2I^{2}\overline{S}\sum_{\mathbf{q}}%
\frac{\widetilde{n}_{\mathbf{k-q}}}{E-t_{\mathbf{k-q}}+\omega _{\mathbf{q}%
}^{{}}}, \nonumber \\
\Sigma _{\mathbf{k\downarrow }}^{{}}(E) &=&2I^{2}\overline{S}\sum_{\mathbf{q}%
}\frac{1-\widetilde{n}_{\mathbf{k-q}}}{E-t_{\mathbf{k-q}}-\omega _{\mathbf{q}%
}^{{}}}  \label{sigfm}
\end{eqnarray}%
The next-leading singular contribution, similarly to (\ref{sig3}) (second
term in the brackets), come from static correlators and are formally reduced
to renormalization of occupation numbers:

\begin{eqnarray}
\widetilde{n}_{\mathbf{k\uparrow }} &=&n_{\mathbf{k\uparrow }}-\frac{1}{%
\overline{S}}\sum_{m^{\prime }\mathbf{q}}\langle c_{\mathbf{k+q}m^{\prime }%
\mathbf{\downarrow }}^{\dagger }c_{\mathbf{k}m\mathbf{\uparrow }}^{{}}S_{%
\mathbf{q}}^{+}\rangle  \label{distr} \\
\widetilde{n}_{\mathbf{k\downarrow }} &=&n_{\mathbf{k\downarrow }}+\frac{1}{%
\overline{S}}\sum_{m^{\prime }\mathbf{q}}\langle c_{\mathbf{k-q}m^{\prime }%
\mathbf{\uparrow }}^{\dagger }c_{\mathbf{k}m\mathbf{\downarrow }}^{{}}S_{-%
\mathbf{q}}^{-}\rangle
\end{eqnarray}%
Calculating the corresponding Green's function yields%
\begin{equation}
\langle \!\langle S_{\mathbf{q}}^{+}|c_{\mathbf{k+q}m^{\prime }\mathbf{%
\downarrow }}^{\dagger }c_{\mathbf{k}m\mathbf{\uparrow }}^{{}}\rangle
\rangle _{\omega }=-\frac{2I\overline{S}}{\omega -\omega _{\mathbf{q}}}\frac{%
n_{\mathbf{k\uparrow }}-n_{\mathbf{k+q\downarrow }}}{\omega +t_{\mathbf{%
k\uparrow }}-t_{\mathbf{k+q\downarrow }}}
\end{equation}%
Using the spectral representation for the retarded Green's function we obtain%
\begin{equation}
\langle c_{\mathbf{k-q}m^{\prime }\mathbf{\uparrow }}^{\dagger }c_{\mathbf{k}%
m\mathbf{\downarrow }}S_{\mathbf{q}}^{-}\rangle =\int_{-\infty }^{0}\omega
d\omega \left( -\frac{\partial n_{\mathbf{k\downarrow }}}{\partial t_{%
\mathbf{k\downarrow }}}\right) \frac{2I\overline{S}}{\omega -\omega _{%
\mathbf{q}}}\delta (\omega +t_{\mathbf{k\uparrow }}-t_{\mathbf{k+q\downarrow
}})
\end{equation}%
the coefficient at the $\delta $-function being just the contribution of the
layer $t_{\mathbf{k+q\downarrow }}-t_{\mathbf{k\uparrow }}=\omega =C$. Note
that the correction to magnon frequency and magnetization can be obtained in
the same manner via magnon damping (cf. Ref. \cite{IKFTT}).

This just gives the singular correction to $\Sigma _{\mathbf{k\downarrow }%
}^{(3)}(E)$. Note that this does not survive in the limit of large $N$. At
the same time, corrections to $\Sigma _{\mathbf{k\uparrow }}^{(3)}(C)$ are
absent since $C=t_{\mathbf{k+q\uparrow }}-t_{\mathbf{k\downarrow }}>0$. When
averaging over $\mathbf{k}$ we have%
\begin{multline}
\delta \langle \Sigma _{\mathbf{k\downarrow }}^{(2)}(C)-\Sigma _{\mathbf{%
k\uparrow }}^{(2)}(C)\rangle /\delta C =-4I^{2}\overline{S}\rho ^{-2}\sum_{%
\mathbf{kk}^{\prime }\mathbf{k}''}\delta (t_{\mathbf{k}})\delta (t_{\mathbf{k}%
^{\prime }})\\
\times \left( \frac{1}{C-\omega _{\mathbf{k-k}^{\prime }}}+\frac{1}{%
C+\omega _{\mathbf{k-k}^{\prime }}}\right)
\end{multline}%
\begin{multline}
\delta \langle \Sigma _{\mathbf{k\downarrow }}^{(3)}(C)\rangle /\delta C
=8MI^{3}\overline{S}\rho ^{-2}\sum_{\mathbf{kk}^{\prime }\mathbf{k}''%
}\delta (t_{\mathbf{k}})\delta (t_{\mathbf{k}^{\prime }})\delta (t_{\mathbf{%
k}''}) \\
\times \frac{C}{(C-\omega _{\mathbf{k-k}^{\prime }})(C+\omega _{\mathbf{k}%
^{\prime }\mathbf{-k}''})}
\end{multline}%
which yields the required cutoffs at the magnon frequency in (\ref{gl}%
).

The correction to the magnon frequency is the same as in the one-loop
consideration \cite{IK}

\begin{equation}
\delta \omega _{\mathbf{q}}/\omega _{\mathbf{q}} =2(1-\widetilde{\alpha }_{%
\mathbf{q}})\delta \overline{S}/S  \label{dwfm}
\end{equation}
\begin{equation}
\widetilde{\alpha }_{\mathbf{q}} =\sum_{\mathbf{R}}J_{\mathbf{R}%
}\left\vert \langle e^{i\mathbf{kR}}\rangle _{t_{\mathbf{k}%
}=E_{F}}^{2}\right\vert ^{2}[1-\cos \mathbf{qR]/}\sum_{\mathbf{R}}J_{\mathbf{%
R}}[1-\cos \mathbf{qR]}
\end{equation}

For an antiferromagnetic structure with the wavevector $\mathbf{Q}$ the
electron spectrum contains the AFM gap $I\overline{S}$,
\begin{equation}
E_{\mathbf{k}}=\frac{1}{2}(t_{\mathbf{k}}+t_{\mathbf{k+Q}})\pm \lbrack \frac{%
1}{4}(t_{\mathbf{k}}-t_{\mathbf{k+Q}})^{2}+I^{2}\overline{S}^{2}]^{1/2}
\label{afgap}
\end{equation}%
The renormalization of $I$ is obtained from the second-order correction to
the anomalous Green's function in the local coordinate system (cf. Ref. \cite%
{afm}),%
\begin{equation}
\sum\limits_{mm^{\prime }}\langle \langle c_{\mathbf{k}m\mathbf{\uparrow }%
}^{{}}|c_{\mathbf{k+Q}m^{\prime }\mathbf{\downarrow }}^{\dagger }\rangle
\rangle _{E}=-M\frac{I\overline{S}-\Sigma _{\mathbf{k,k+Q}}^{AFM}(E)}{(E-t_{%
\mathbf{k}})(E-t_{\mathbf{k+Q}})},
\end{equation}%
so that
\begin{equation}
\delta I_{ef}=-\Sigma _{\mathbf{k,k+Q}}^{AFM}(E)/\overline{S}.
\end{equation}%
The calculation of the off-diagonal self-energy gives (we consider for
simplicity a two-sublattice situation with $\omega _{\mathbf{q}}^{{}}=\omega
_{\mathbf{q+Q}}^{{}}$)%
\begin{multline}
\Sigma _{\mathbf{k,k+Q}}^{AFM}(E) =\frac{1}{4}I^{2}\sum_{m\mathbf{q}%
}\langle \!\langle (S_{\mathbf{q}}^{+}+S_{\mathbf{q}}^{-})c_{\mathbf{k-q}m%
\mathbf{\uparrow }}^{{}}\\
+(S_{\mathbf{q}}^{+}-S_{\mathbf{q}}^{-})c_{\mathbf{k-q-Q}m\mathbf{\downarrow }}^{{}}\\
-2\delta S_{\mathbf{q}}^{z}c_{\mathbf{k-q-Q}m\mathbf{\downarrow }%
}^{{}}|\\
-c_{\mathbf{k+p-Q}m^{\prime }\mathbf{\downarrow }}^{\dagger }(S_{%
\mathbf{p}}^{+}+S_{\mathbf{p}}^{-})
+c_{\mathbf{k+p}m^{\prime }\mathbf{%
\uparrow }}^{\dagger }(S_{\mathbf{p}}^{+}-S_{\mathbf{p}}^{-})-2c_{\mathbf{k+p%
}m^{\prime }\mathbf{\uparrow }}^{\dagger }\delta S_{\mathbf{p}}^{z}\rangle
\!\rangle _{E}.
\end{multline}%
We derive%
\begin{equation}
\Sigma _{\mathbf{k,k+Q}}^{AFM}(E)=2I^{2}S\sum_{\mathbf{q}}\frac{(E-t_{%
\mathbf{k-q}})\widetilde{n}_{\mathbf{k-q}}}{(E-t_{\mathbf{k-q}})^{2}-\omega
_{\mathbf{q}}^{2}}  \label{sigafm}
\end{equation}%
\begin{multline}
\widetilde{n}_{\mathbf{k}}=n_{\mathbf{k}}-\frac{1}{2S}\sum_{m^{\prime }%
\mathbf{q}}\langle c_{\mathbf{k+q}m^{\prime }\mathbf{\uparrow }}^{\dagger
}c_{\mathbf{k}m\mathbf{\uparrow }}^{{}}(S_{\mathbf{q}}^{+}-S_{\mathbf{q}%
}^{-})\\
-c_{\mathbf{k}m^{\prime }\mathbf{\uparrow }}^{\dagger }c_{\mathbf{k+q-Q%
}m\mathbf{\downarrow }}^{{}}(S_{\mathbf{q}}^{+}+S_{\mathbf{q}}^{-})\rangle
\end{multline}%
The Green's function needed is calculated as
\begin{multline}
-\sum_{m^{\prime }\mathbf{q}}\langle \!\langle S_{\mathbf{q}}^{+}+S_{\mathbf{%
q}}^{-}|c_{\mathbf{k+q-Q}m^{\prime }\mathbf{\downarrow }}^{\dagger }c_{%
\mathbf{km\uparrow }}^{{}}\rangle \!\rangle _{\omega }\\
=\sum_{m^{\prime }%
\mathbf{q}}\langle \!\langle S_{\mathbf{q}}^{+}-S_{\mathbf{q}}^{-}|c_{%
\mathbf{k+q}m^{\prime }\mathbf{\uparrow }}^{\dagger }c_{\mathbf{km\uparrow }%
}^{{}}\rangle \!\rangle _{\omega }\\
=2MI\overline{S}\sum_{\mathbf{q}}\frac{%
\omega }{\omega _{{}}^{2}-\omega _{\mathbf{q}}^{2}}\frac{n_{\mathbf{k}}-n_{%
\mathbf{k+q}}}{\omega -t_{\mathbf{k+q}}+t_{\mathbf{k}}}
\end{multline}%
so that we obtain%
\begin{equation}
\delta \langle \Sigma _{\mathbf{k,k+Q}}^{(2)}(C)\rangle /\delta C =2I^{2}%
\overline{S}\sum_{\mathbf{kk}^{\prime }\mathbf{k}''}\delta (t_{\mathbf{k}%
})\delta (t_{\mathbf{k}^{\prime }})\frac{C^{{}}}{C^{2}-\omega _{\mathbf{k-k}%
^{\prime }}^{2}}
\end{equation}%
\begin{multline}
\delta \langle \Sigma _{\mathbf{k,k+Q}}^{(3)}(C)\rangle /\delta C =-2MI^{3}%
\overline{S}S\rho ^{-2}\\
\times \sum_{\mathbf{kk}^{\prime }\mathbf{k}''}\delta (t_{%
\mathbf{k}})\delta (t_{\mathbf{k}^{\prime }})\delta (t_{\mathbf{k}''})\frac{%
C^{3}}{(C^{2}-\omega _{\mathbf{k-k}^{\prime }}^{2})(C^{2}-\omega _{\mathbf{k}%
^{\prime }\mathbf{-k}''}^{2})}
\end{multline}%
in agreement with (\ref{sef}).

For the staggered AFM ordering in a cubic lattice with the dimensionality $%
d$ one has \cite{731}%
\begin{eqnarray}
\delta \overline{\omega }/\overline{\omega } &=&(1-\alpha ^{\prime })\delta
\overline{S}/S, \label{alph} \\
\alpha ^{\prime } &\simeq &2(d-1)\frac{J_{2}}{J_{1}}\left\vert \left\langle
\exp (i\mathbf{kR}_{2})\right\rangle _{t_{\mathbf{k}}=0}\right\vert ^{2}
\end{eqnarray}%
where $J_{1}$ and $J_{2}$ are the exchange integrals between nearest and
next-nearest neighbors ($|J_{1}|\gg |J_{2}|$), $\mathbf{R}_{2}$ runs over
the next-nearest neighbors.

\section*{Appendix C. Scaling functions}

For the paramagnetic phase we have
\begin{equation}
\eta ^{PM}(\frac{\overline{\omega }}{C})=\operatorname{Re}\int_{-\infty }^{\infty
}d\omega \langle \mathcal{J}_{\mathbf{k-k}^{\prime }}(\omega )\rangle
_{t_{k}=t_{k^{\prime }}=E_{F}}\frac{1}{1-(\omega +i0)/C}
\end{equation}%
In the spin-diffusion approximation (\ref{JD}) we obtain
\begin{equation}
\eta ^{PM}(\frac{\overline{\omega }}{C})=\left\langle \frac{1}{1+\mathcal{D}(%
\mathbf{k-k}^{\prime })^{2}/C^{2}}\right\rangle _{t_{k}=t_{k^{\prime }}=0}
\end{equation}%
where $\overline{\omega }=4\mathcal{D}k_{F}^{2},$ the averages go over the
Fermi surface. Integration yields%
\begin{equation*}
\eta ^{PM}(x)=\left\{
\begin{array}{cc}
\arctan x/x & d=3 \\
\{\frac{1}{2}[1+(1+x^{2})^{1/2}]/(1+x^{2})\}^{1/2} & d=2%
\end{array}%
\right.
\end{equation*}%
In the FM and AFM phases for simple magnetic structures we have
\begin{equation}
\eta \left( \overline{\omega }_{ef}/|C|,\delta \right) =\operatorname{Re}%
\left\langle \left( 1-(\omega _{\mathbf{k-k}^{\prime }}^{{}}+i\delta
)^{2}/C^{2}\right) ^{-1}\right\rangle _{t_{k}=t_{k^{\prime }}=0} \label{eta11}
\end{equation}%
where $\delta $ is a cutoff owing to damping. (Note that in the FM case with
$N>2$ this expression should be generalized since spin-up and spin-down
contributions are asymmetric \cite{IK}.) For an isotropic 3D ferromagnet
integration in (\ref{eta11}) for quadratic spin-wave spectrum $\omega _{%
\mathbf{q}}\propto q^{2}$ yields
\begin{equation}
\eta ^{FM}(x)=\left\{
\begin{array}{cc}
\frac{1}{4x}\ln \{[(1+x)^{2}+\delta ^{2}]/[(1-x)^{2}+\delta ^{2}]\} & d=3 \\
\frac{1}{2}\operatorname{Re}[(1-i\delta -x)^{-1/2}+(1+i\delta +x)^{-1/2}] & d=2%
\end{array}%
\right.  \label{intfm}
\end{equation}%
where $x=\overline{\omega }_{ef}/|C|,$ $\overline{\omega }=\omega _{2k_{F}}.$

For an antiferromagnet integration  with the linear spin-wave spectrum $%
\omega _{\mathbf{q}}\propto q$ gives
\begin{equation}
\eta ^{AFM}(x,z)=\left\{
\begin{array}{cc}
-(2x^{2})^{-1} \ln [(1-x^{2})^{2}+4\delta ^{2}] & d=3 \\
\operatorname{Im}[x^{2}-(1+i\delta )^{2}]^{-1/2} & d=2%
\end{array}%
\right.  \label{intafm}
\end{equation}%
Thus $\Psi $ becomes bounded from above:
\begin{equation}
\Psi _{\max }=\eta _{\max }\simeq \eta (1)\simeq \left\{
\begin{tabular}{ll}
$\frac{1}{2}\ln \delta $ & 3D FM \\
$\ln \delta $ & 3D AFM \\
$\frac{1}{2^{3/2}}\delta ^{-1/2}$ & 2D FM \\
$\frac{1}{2}\delta ^{-1/2}$ & 2D AFM%
\end{tabular}%
\right.  \label{delta}
\end{equation}
One can see that the scaling
functions for the ordered phases contain Van Hove  singularities at $x=1.$ Presence of
such singularities is a general property which does not depend on the spectrum
model. The function $\eta ^{AFM}(x)$ ($%
d=3)$ changes its sign at $x=\sqrt{2}$. For $d=2$, 
 $\eta ^{AFM}(x)$ vanishes discontinuously at $x>1$, but a smooth
contribution occurs for more realistic models of magnon spectrum.

A more detailed analysis of the scaling function singularities is presented
in Ref.\cite{731}.

\end{document}